\documentclass[aps,preprint,amsmath,amssymb,superscriptaddress]{revtex4}

\usepackage{graphicx}
\begin{document}

\title{Probe of extra dimensions in $\gamma q\to \gamma q$ at the LHC}

\author{\.{I}. \c{S}ahin}
\email[]{inancsahin@ankara.edu.tr}
 \affiliation{Department of
Physics, Faculty of Sciences, Ankara University, 06100 Tandogan,
Ankara, Turkey}

\author{A. A. Billur}
\email[]{abillur@cumhuriyet.edu.tr} \affiliation{Department of
Physics, Cumhuriyet University, 58140 Sivas, Turkey}

\author{S. C. \.{I}nan}
\email[]{sceminan@cumhuriyet.edu.tr}  \affiliation{Department of
Physics, Cumhuriyet University, 58140 Sivas, Turkey}

\author{B. \c{S}ahin}
\email[]{banusahin@ankara.edu.tr} \affiliation{Department of
Physics, Faculty of Sciences, Ankara University, 06100 Tandogan,
Ankara, Turkey}

\author{M. K\"{o}ksal}
\email[]{mkoksal@cumhuriyet.edu.tr} \affiliation{Department of
Physics, Cumhuriyet University, 58140 Sivas, Turkey}

\author{P. Tekta\c{s}}
 \affiliation{Department of Physics, Bulent Ecevit University, 67100 Zonguldak, Turkey}

\author{E. Al{\i}c{\i}}
 \affiliation{Department of Physics, Bulent Ecevit University, 67100 Zonguldak, Turkey}

\author{R. Y{\i}ld{\i}r{\i}m}
  \affiliation{Department of
Physics, Cumhuriyet University, 58140 Sivas, Turkey}

\begin{abstract}
We have examined TeV scale effects of extra spatial dimensions
through the processes $\gamma q\to \gamma q$ where $q=u,d,c,s,b,\bar
u,\bar d, \bar c, \bar s, \bar b$. These processes have been treated
in a photon-proton collision via the main reaction $pp\to p\gamma
p\to p\gamma qX$ at the LHC. We have employed equivalent photon
approximation for incoming photon beams and performed statistical
analysis for various forward detector acceptances.
\end{abstract}

\pacs{bbbb}

\maketitle

\section{Introduction}

Extra spatial dimensions that show themselves near the TeV scale
have been widely studied in particle physics since the pioneering
works of Arkani-Hamed, Dimopoulos and Dvali (ADD)
\cite{ArkaniHamed:1998rs,ArkaniHamed:1998nn,Antoniadis:1998ig}. Soon
after the work of ADD a warped model was proposed by Randall and
Sundrum (RS) \cite{Randall:1999ee}. According to ADD and RS models,
extra spatial dimensions can have observable effects at the TeV
scale physics. The possibility of these extra dimensions has been
probed in the past colliders but no evidence has been found. The
Large Hadron Collider (LHC) offers the opportunity of a very rich
physics program. Signals confirming the existence of extra
dimensions might become detectable in the high energetic collisions
of the LHC. Phenomenological studies on extra dimensions involving
quark-quark, gluon-gluon or quark-gluon collisions at the LHC are
widespread in the literature. On the other hand, extra dimensions
have been much less studied in photon-induced reactions
($\gamma\gamma$ or $\gamma$-proton collisions) at the LHC.

In an usual proton-proton deep inelastic scattering (DIS) processes
both of the incoming protons dissociate into partons. Due to proton
remnants, usual DIS processes do not provide a clean environment.
Jets coming from proton remnants create some uncertainties and make
it difficult to discern the signals which may arise from the new
physics beyond the standard model. On the other hand, in
$\gamma\gamma$ or $\gamma$-proton collisions with quasireal photons,
photon emitting protons remains intact. $\gamma\gamma$ processes
provide the most clean channels due to absence of the remnants of
both proton beams. Whereas in $\gamma$-proton processes one of the
incoming protons dissociates into partons but other proton remains
intact. Midway from proton-proton DIS to $\gamma\gamma$,
$\gamma$-proton processes have less experimental uncertainties
compared with proton-proton processes. Furthermore, they have higher
energy reach and effective luminosity with respect to $\gamma\gamma$
processes \cite{rouby,deFavereaudeJeneret:2009db,Schul:2011xwa}.

\begin{figure}
\includegraphics[scale=0.5]{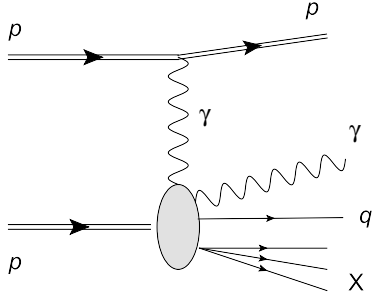}
\caption{Schematic diagram for the reaction $p p\to p\gamma p\to p
\gamma q X$. \label{fig1}}
\end{figure}

In this work we have investigated TeV scale effects of extra
spatial dimensions via the process $\gamma q\to \gamma q$ at the
LHC. The process $\gamma q\to \gamma q$ takes part as a subprocess
in the main reaction $pp\to p\gamma p\to p\gamma qX$ (Fig.\ref
{fig1}). The photon which enters the subprocess is emitted from
one of the proton beams and described by equivalent photon
approximation (EPA) \cite{budnev1975,baur2002,piotrzkowski2001}.
In the framework of EPA, virtuality of the quasireal photons is
very low. Hence when a proton emits a quasireal photon, it does
not dissociate into partons. In EPA, quasireal photons carry a
small transverse momentum. Therefore photon emitting intact
protons deviate slightly from their trajectory along the beam
path. They are generally scattered with very small angles from the
beam pipe and exit the central detector without being detected.
Consequently, detection of intact protons needs forward detector
equipment in addition to central detectors. It is foreseen to
equip ATLAS and CMS central detectors with very forward detectors
which can detect intact scattered protons with a very large
pseudorapidity. A project called AFP (ATLAS Forward Physics) that
aims to install very forward detectors located at distances 220
and 420 m from the interaction point, is under evaluation in the
ATLAS collaboration \cite{royon2007,albrow2009}. The acceptance
proposed by AFP project is $0.0015<\xi<0.15$ where $\xi$ is the
momentum fraction loss of the intact scattered protons.
Mathematically speaking, it is defined by the formula
$\xi=(|\vec{p}|-|\vec{p}^{\,\,\prime}|)/|\vec{p}|$. Here $\vec{p}$
is the momentum of incoming proton and $\vec{p}^{\,\,\prime}$ is
the momentum of intact scattered proton. At the LHC energies, it
is a good approximation to write $\xi=\frac{E_\gamma}{E}$ where
$E_\gamma$ is the energy of the emitted quasireal photon and $E$
is the energy of the incoming proton. There are also other
scenarios with different acceptances. When forward detectors are
placed closer to the interaction point they can detect protons
with higher $\xi$. In the CMS-TOTEM forward detector scenario, a
forward detector acceptance of $0.0015<\xi<0.5$ is considered
\cite{avati2006,kepka2008}. This wide acceptance range is provided
by the use of the detectors of TOTEM experiment at 147 and 220 m
from the CMS interaction point in addition to forward detectors at
420 m.

Existence of photon-induced reactions in a hadron collider is not
merely a theoretical hypothesis. Photon-induced reactions in a
hadron-hadron collision were verified experimentally at the
Fermilab Tevatron \cite{cdf1,cdf2,cdf3}. The reactions such as $p
\bar p\to p \gamma \gamma \bar p\to p e^+ e^- \bar p$
\cite{cdf1,cdf2}, $p \bar p\to p \gamma \gamma \bar p\to p\; \mu^+
\mu^- \bar p$ \cite{cdf2,cdf3}, $p \bar p\to p \gamma \bar p\to
p\; J/\psi\;(\psi(2S)) \bar p$ \cite{cdf3} were observed by the
CDF collaboration. From the early LHC data obtained in
proton-proton collisions at $\sqrt s=7$ TeV, two-photon reactions
$p p\to p \gamma \gamma p\to p \mu^+ \mu^- p$ and $p p\to p \gamma
\gamma p\to p e^+ e^- p$ have been observed by the CMS
Collaboration \cite{Chatrchyan:2011ci,Chatrchyan:2012tv}. Probing
new physics via photon-photon and photon-proton reactions at the
LHC  has been studied in the literature. Phenomenological studies
cover a wide range of new physics such as supersymmetry, extra
dimensions, unparticle physics, anomalous interaction of standard
model particles, magnetic monopoles, etc.
\cite{deFavereaudeJeneret:2009db,kepka2008,Ginzburg:1998vb,Ginzburg:1999ej,fd12,fd13,lhc1,Pierzchala2008,lhc2,lhc3,lhc4,
Dougall:2007tt,Chaichian:2009ts,Piotrzkowski:2009sa,lhc5,Goncalves:2010dw,lhc6,lhc7,lhc8,lhc9,lhc10,lhc11,lhc12,lhc13,
Epele:2012jn,lhc14,lhc15,lhc16,Senol:2013ym}.

In this paper we aim to constrain model parameters of ADD and RS
models in a {\it quasireal photon-proton} deep inelastic scattering
process. As far as we know, ADD or RS model of extra dimensions has
not been studied and model parameters have not been constrained in
any phenomenological study in the context of {\it quasireal
photon-proton} deep inelastic scattering at the LHC. The subprocess
$\gamma q\to \gamma q$ that we have considered, is the simplest
process which appears in a photon-proton collision. It is very
similar to Compton scattering which is one of the fundamental
processes in particle physics. $\gamma q\to \gamma q$ may take part
as a subprocess in any reaction where the electromagnetic
interaction of quarks is considered. Hence, it is important to know
the effect of new physics coming from extra dimensional theories to
this particular process.

\section{ADD Model of Large Extra Dimensions}

ADD model was proposed as a solution to the hierarchy problem
which is known as the unexplained large difference between the
electroweak scale $\sim {\cal O}(100\; \text{GeV})$ and the Planck
scale $M_{Pl}\sim {\cal O}(10^{19}\; \text{GeV})$
\cite{ArkaniHamed:1998rs,ArkaniHamed:1998nn,Antoniadis:1998ig}.
According to ADD model, gravity can propagate in a $4+\delta$
dimensional space called "bulk" but Standard Model (SM) particles
are confined in a hypersurface called "brane". Using Gauss' law in
arbitrary dimensions, 4-dimensional Planck scale can be related to
the $(4+\delta)$-dimensional fundamental scale $M_D$ through the
formula
\begin{eqnarray}
\label{reducedplanck} M_{Pl}^2=8\pi R^\delta M_D^{2+\delta}.
\end{eqnarray}
Here, $R$ is the radius of the compactified extra dimensional space
of dimension $\delta$ and volume $V_\delta=(2\pi R)^\delta$. In the
ADD model, the hierarchy is eliminated by choosing the
compactification radius large. For instance if we choose $R\sim 0.1$
mm for $\delta=2$ then $M_D$ is at the order of ${\cal O}(1\; TeV)$.
An important consequence of large extra dimensions is the tower of
Kaluza-Klein (KK) modes. Solutions of linear Einstein equations in
$4+\delta$ dimension manifest themselves as a set of states
separated in mass by ${\cal O}(\frac{1}{R})$ in 4 dimension
\cite{Giudice:1998ck,Han:1998sg}. In 4-dimensional effective theory
we have spin-2 and spin-0 KK states that can interact with SM fields
on the brane. Spin-2 and spin-0 states are sometimes called
KK-gravitons and gravitational scalars respectively. KK-gravitons
couple to the energy-momentum tensor of the SM fields. Although
their coupling to SM fields is suppressed by a factor proportional
to $\frac {1} {M_{Pl}}$, summation of enormous number of KK states
in a tower provides an effective coupling of order $\frac{1}{M_D}$.
Therefore, KK-gravitons can have observable effects at the TeV
scale. Gravitational scalars are coupled only to the trace of the
energy-momentum tensor. Since the trace of the energy-momentum
tensor is zero for massless particles, the coupling of gravitational
scalar to photons is zero at the tree-level. Hence, we will neglect
gravitational scalars during amplitude calculations. Feynman rules
for KK-gravitons were given in \cite{Giudice:1998ck,Han:1998sg}.

\begin{figure}
\includegraphics[scale=1]{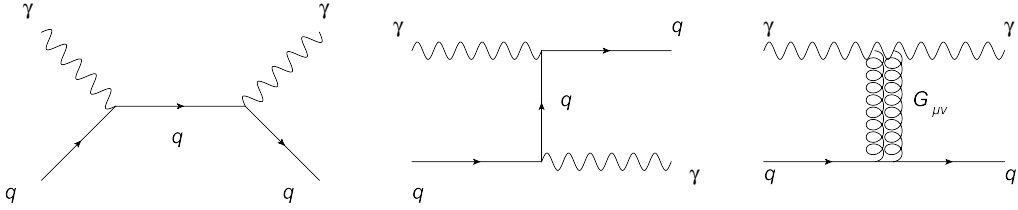}
\caption{Tree-level Feynman diagrams for the subprocess $\gamma q
\to \gamma q$ ($q=u,d,c,s,b,\bar u,\bar d, \bar c, \bar s, \bar
b$) in the presence of Kaluza-Klein graviton
mediation.\label{fig2}}
\end{figure}

The process $\gamma q\to \gamma q$ is described by 3 tree-level
diagrams (Fig.\ref{fig2}). The polarization summed amplitude square
can be written as
\begin{eqnarray}
\label{amplitude} |M|^2=|M_{SM}|^2+|M_{KK}|^2+|M_{int}|^2
\end{eqnarray}
where $M_{SM}$ is the SM amplitude, $M_{KK}$ is the amplitude for
the t-channel KK-graviton exchange and $|M_{int}|^2$ represents
interference terms between the SM and the KK amplitudes. Analytical
expressions for SM, KK and interference terms as a function of
Mandelstam parameters $\hat s$, $\hat t$ and $\hat u$ are
\begin{eqnarray}
\label{amplitudeSM} |M_{SM}|^2=-8g_e^4q^4 &&\left[ \frac{1}{(\hat
s-m_q^2)^2}[3m_q^4+\hat s \hat u-m_q^2(5\hat s+2\hat t+3\hat u)]
\right. \nonumber
\\ && \left. +\frac{1}{(\hat u-m_q^2)^2}[3m_q^4+\hat s \hat u-m_q^2(3\hat s+2\hat t+5\hat u)] \right. \nonumber
\\ && \left.-\frac{2}{(\hat s-m_q^2)(\hat u-m_q^2)} [6m_q^4+2m_q^2\hat t-m_q^2(\hat s+4\hat t+\hat u)] \right]
\end{eqnarray}

\begin{eqnarray}
\label{amplitudeKK}
 |M_{KK}|^2=\frac{1}{2\bar {M_{Pl}}^4}|D(\hat t)|^2\left[(4m_q^2-\hat
 t)\hat t+(\hat s-\hat u)^2\right]\left[2m_q^4+\hat s^2+\hat u^2-2m_q^2(\hat s-\hat t+\hat u)\right]
\end{eqnarray}

\begin{eqnarray}
\label{amplitudeint}
 |M_{int}|^2=\frac {g_e^2q^2(D(\hat t)+D^\ast(\hat t))}{2\bar {M_{Pl}}^2}&&\left\{ \frac{1}{(\hat
s-m_q^2)}[8m_q^6-2m_q^4(\hat s-5\hat t+7\hat u)\right. \nonumber
\\ &&\left.+(\hat s- \hat t-\hat u)(3\hat s^2+2\hat s\hat t+ \hat t^2-(\hat s+\hat t)\hat u)\right. \nonumber \\ &&\left.
+m_q^2(-3\hat s^2+10\hat s \hat t+\hat t^2+6\hat s \hat u-2\hat
t\hat u+5\hat u^2)] \right. \nonumber \\ &&\left.+(\hat s
\longleftrightarrow \hat u) \right\}
\end{eqnarray}
where $g_e=\sqrt{4\pi\alpha}$ and $m_q$ is the mass of the quark.
$q$ is the quark charge which is given in units of positron
charge. In eqs.(\ref{amplitudeSM}-\ref{amplitudeint}) we do not
write the factor due to initial spin average. $\bar
M_{Pl}=M_{Pl}/\sqrt{8\pi}$ is the reduced Planck mass. $D(\hat t)$
denotes propagator factors which are summed over infinite tower of
KK modes. The existence of this infinite sum creates ultraviolet
divergences even in tree-level processes. We employ the cutoff
procedure that was assumed in Ref.\cite{Giudice:1998ck} for
phenomenological applications:
\begin{eqnarray}
\label{cutoff} \frac{1}{\bar {M_{Pl}}^2}D(\hat t)=\frac{1}{\bar
{M_{Pl}}^2}\sum_n\frac{1}{t-m_n^2}\equiv\frac{4\pi}{\Lambda_T^4}\;\;\;\;\;\;\;\;\;\text{for}\;\;
\delta>2
\end{eqnarray}
Here, $\Lambda_T$ is an effective cutoff scale. Its dependence on
$M_D$ can be identified with some knowledge of the underlying
quantum gravity theory. In case of string theory, the inequality
$M_D>\Lambda_T$ can be written \cite{Giudice:1998ck}. As a
consequence, any lower bound for $\Lambda_T$ also serves as a lower
bound for $M_D$.

The cross section for the main process $pp\to p\gamma p\to p\gamma
qX$ can be obtained by integrating the cross section for the
subprocess $\gamma q\to \gamma q$ over the photon and quark
distributions:
\begin{eqnarray}
\label{mainprocess}
 \sigma\left(p p\to p\gamma p\to p \gamma q
X\right)=\sum_q\int_{\xi_{min}}^{\xi_{max}} {dx_1 }\int_{0}^{1}
{dx_2}\left(\frac{dN_\gamma}{dx_1}\right)\left(\frac{dN_q}{dx_2}\right)
\hat{\sigma}_{\gamma q\to \gamma q}(\hat s)
\end{eqnarray}
where $x_1$ is the fraction which represents the ratio between the
scattered equivalent photon and initial proton energy and $x_2$ is
the momentum fraction of the proton's momentum carried by the
struck quark. $\frac{dN_\gamma}{dx_1}$ and $\frac{dN_q}{dx_2}$ are
the equivalent photon and quark distribution functions. Analytical
expression for $\frac{dN_\gamma}{dx_1}$ is given in the Appendix.
Quark distribution functions have been evaluated numerically by
using a code MSTW2008 \cite{Martin:2009iq}.  The summation in
(\ref{mainprocess}) is performed over the following subprocesses:
\begin{eqnarray}
\label{subprocesses} &&\text{(i)}\;\;\gamma u \to \gamma u
\;\;\;\;\;\;\;\;\;\;\;\;\text{(vi)}\;\;\gamma \bar u \to \gamma
\bar u  \nonumber
\\ &&\text{(ii)}\;\;\gamma d \to \gamma d
\;\;\;\;\;\;\;\;\;\;\;\;\text{(vii)}\;\;\gamma \bar d \to \gamma
\bar d \nonumber
\\&&\text{(iii)}\;\;\gamma c \to \gamma c
\;\;\;\;\;\;\;\;\;\;\text{(viii)}\;\;\gamma \bar c \to \gamma \bar c  \\
&&\text{(iv)}\;\;\gamma s \to \gamma s
\;\;\;\;\;\;\;\;\;\;\;\text{(ix)}\;\;\gamma \bar s \to \gamma \bar
s  \nonumber \\&&\text{(v)}\;\;\gamma b \to \gamma b
\;\;\;\;\;\;\;\;\;\;\;\;\;\text{(x)}\;\;\gamma \bar b \to \gamma
\bar b  \nonumber
\end{eqnarray}
During all calculations in this paper we assume that center-of-mass
energy of the proton-proton system is 14 TeV.

We estimate the sensitivity of the reaction $pp\to p\gamma p\to
p\gamma qX$ to extra dimensions using a simple one parameter
$\chi^2$ criterion without a systematic error. The $\chi^2$ function
is given by
\begin{eqnarray}
\chi^{2}=\left(\frac{\sigma-\sigma_{SM}}{\sigma_{SM} \,\,
\delta}\right)^{2}
\end{eqnarray}
where $\sigma$ is the cross section containing both SM and KK
contributions, $\sigma_{SM}$ is the SM cross section and
$\delta=\frac{1}{\sqrt{N}}$ is the statistical error. Cross sections
used in the $\chi^2$ function are integrated total cross sections
which are defined by Eq. (\ref{mainprocess}). Hence, contributions
from all subprocesses in (\ref{subprocesses}) have been taken into
account. During statistical analysis, the expected number of events
is calculated through the formula: $N=S\times E\times \sigma_{SM}
\times L_{int}$, where, $L_{int}$ is the integrated luminosity, $E$
is the jet reconstruction efficiency and $S$ is the survival
probability factor. We consider a survival probability factor of
$S=0.7$ and jet reconstruction efficiency of $E=0.6$. We have also
imposed a pseudorapidity cut of $|\eta|<2.5$ for final (anti-)quarks
and photons from subprocesses in (\ref{subprocesses}). We have
obtained 95\% confidence level (C.L.) lower bounds for $\Lambda_T$
considering forward detector acceptances of $0.0015<\xi<0.15$,
$0.0015<\xi<0.5$, $0.1<\xi<0.15$ and $0.1<\xi<0.5$. The first two
are the AFP and CMS-TOTEM acceptances as we have mentioned in the
introduction. The last two are the subintervals of the whole AFP and
CMS-TOTEM acceptance regions. Forward detectors have a capability to
detect intact protons in a continuous range of momentum fraction
loss $\xi$. Hence, we can impose cuts and choose to work in a
subinterval of the whole acceptance region. Since the KK terms in
the amplitude square have a higher momentum dependence than the SM
terms, imposing such cuts and removing low energy region of the
whole acceptance range considerably suppress the SM contribution
without minimizing the KK effects.

\begin{figure}
\includegraphics[scale=1]{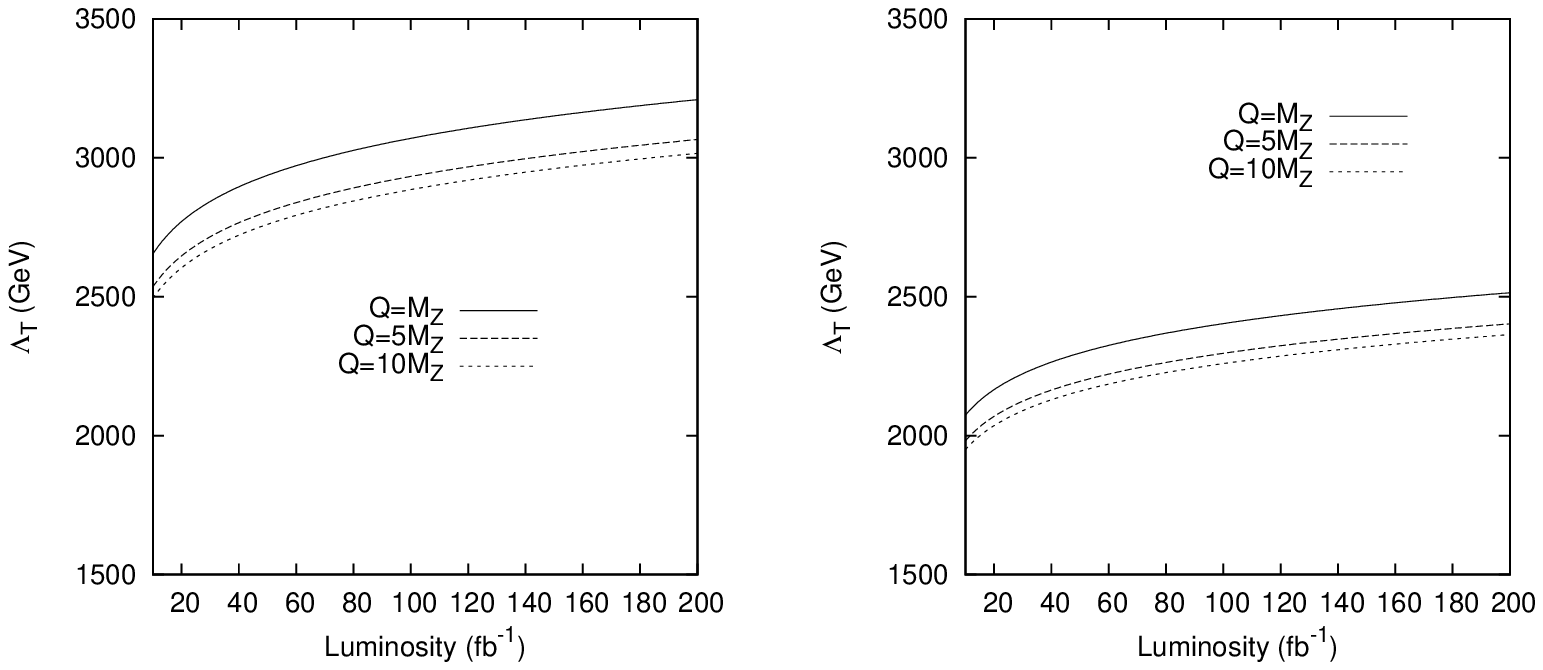}
\caption{95\% C.L. lower bounds for $\Lambda_T$ as a function of
integrated LHC luminosity for forward detector acceptance regions
$0.0015<\xi<0.5$ (left panel) and $0.0015<\xi<0.15$ (right panel).
Legends are for various values of the virtuality of the deep
inelastic scattering. \label{fig3}}
\end{figure}

In Fig.\ref{fig3} we present 95\% C.L. lower bounds for $\Lambda_T$
as a function of integrated LHC luminosity for AFP and CMS-TOTEM
acceptances. At the LHC energies, deep inelastic scattering
processes generally have a very high virtuality. Due to Bjorken
scaling it is expected that quark distribution functions do not
depend significantly on $Q^2$ but only on $x$. Hence, during
numerical calculations a fix $Q^2$ value can be used. Bjorken
scaling is, however, not exact. For this reason, the bounds have
been obtained by considering three different virtualities
$Q^2=M_Z^2,{(5M_Z)}^2$ and $({10M_Z})^2$ for the deep inelastic
scattering where $M_Z$ is the mass of the Z boson. Here, $M_Z$
represents only a scale which is roughly at the order of Standard
Model energies. $Q= M_Z,5M_Z$ and $10M_Z$ are plausible scales for
our process. For instance, if the center-of-mass energy of the
subprocess $\gamma q\to \gamma q$ is $\sqrt{\hat s}=180$ GeV and
outgoing photon is scattered with an angle of $60$ degree at the
center-of-mass system of the incoming photon and quark then the
square of the momentum transferred to the proton is $q^2=- (90
GeV)^2$. Similarly for $\sqrt{\hat s}=900$ GeV and $\sqrt{\hat
s}=1800$ GeV, the corresponding momentum squares are $q^2=- (450
GeV)^2$ and $q^2=- (900 GeV)^2$ respectively. In the laboratory
system, incoming photon and quark do not have fix energies. Instead,
their energies are described by photon and quark distributions. If
we assume that center-of-mass energy of the proton-proton system is
$\sqrt{s}=14$ TeV and upper bound for the forward detector
acceptance is 0.5 (0.15) then the center-of-mass energy of the
subprocess extends up to energies of approximately 9900 GeV (5422
GeV). Hence, it is probable for our subprocess to possesses a
virtuality of $Q^2=M_Z^2,{(5M_Z)}^2$ or $({10M_Z})^2$.

\begin{figure}
\includegraphics[scale=1]{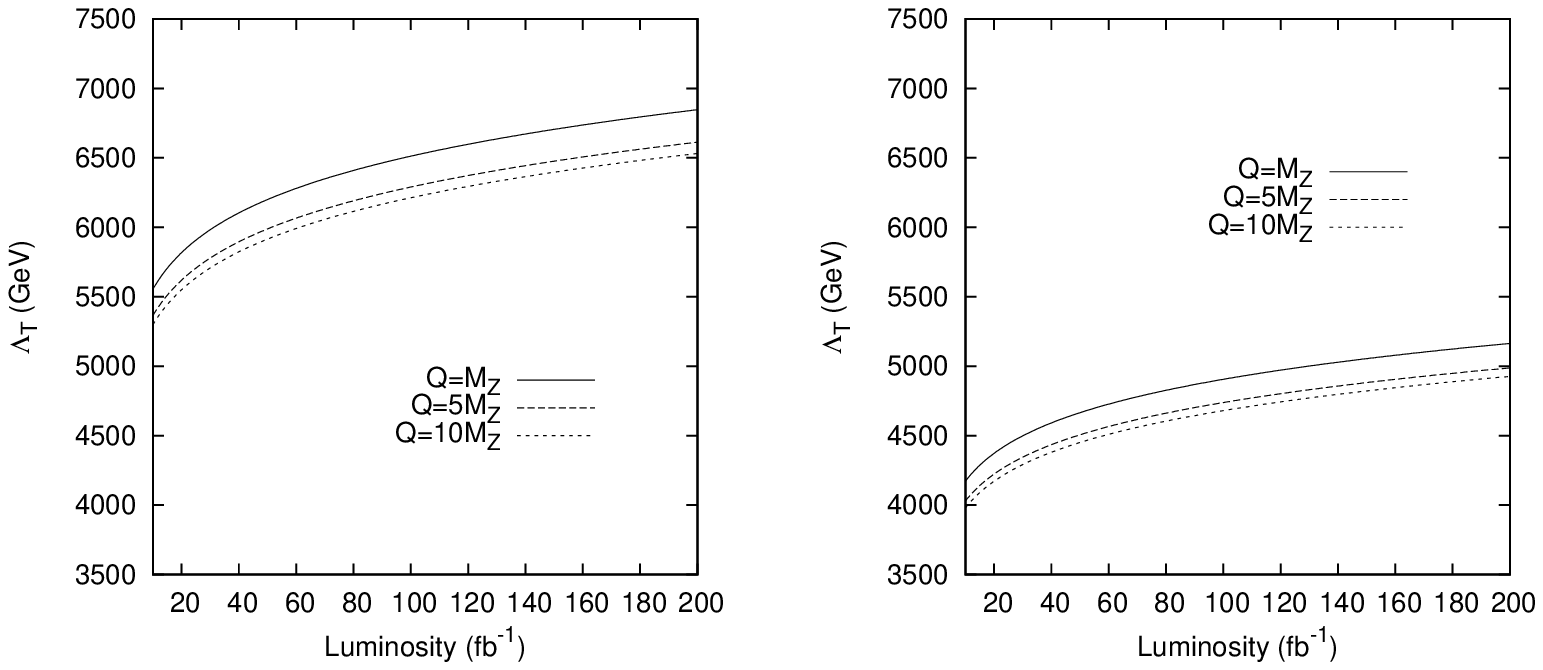}
\caption{95\% C.L. lower bounds for $\Lambda_T$ as a function of
integrated LHC luminosity for forward detector acceptance regions
$0.1<\xi<0.5$ (left panel) and $0.1<\xi<0.15$ (right panel).
Legends are for various values of the virtuality of the deep
inelastic scattering. \label{fig4}}
\end{figure}

In Fig.\ref{fig4} we present the lower bounds for $0.1<\xi<0.15$ and
$0.1<\xi<0.5$ subintervals of the whole AFP and CMS-TOTEM
acceptances. We see from these figures that the bounds for
$0.1<\xi<0.15$ and $0.1<\xi<0.5$ cases are approximately 2 times
stronger than the bounds for $0.0015<\xi<0.15$ and $0.0015<\xi<0.5$
respectively. We have exhibited in Fig.\ref{fig3} and Fig.\ref{fig4}
that limits vary a little with $Q^2$. But the variation in limits is
minor. To be precise, we see from the left panel of Fig.\ref{fig3}
that limits vary approximately by a factor of 1.06 when the square
root of the virtuality $Q$ varies by a factor of 10. (or
equivalently, virtuality $Q^2$ varies by a factor of 100.)

\section{RS Model of Warped Extra Dimensions}

Although the ADD model eliminates the hierarchy between the
electroweak scale and the Planck scale, it introduces a new
hierarchy between the electroweak scale and $\frac{1}{R}$. In this
respect, RS model solves the hierarchy problem without generating
another large hierarchy. In the simplest RS model, we have only one
extra spatial dimension and two branes located at orbifold fixed
points $y=0$ and $y=\pi r_c$ \cite{Randall:1999ee}. Here, $y$
represents the extra dimensional coordinate and $r_c$ is the
compactification radius of the extra dimension. The brane which is
located at $y=\pi r_c$ is called the TeV-brane where SM fields live
on. The brane at $y=0$ is called the Planck-brane. As in the case of
ADD model, gravity can propagate to everywhere. TeV and Planck
branes have different vacuum energies and the 5 dimensional bulk
bounded by these branes has a cosmological constant $\Lambda$.
Assuming four dimensional Poincare invariance, the solution to
Einstein's field equations is given by the following metric
\cite{Randall:1999ee}:
\begin{eqnarray}
\label{RSmetric}
ds^{2}=e^{-2k|y|}\eta_{\mu\nu}dx^{\mu}dx^{\nu}-dy^{2},
\end{eqnarray}
where $k$ is a constant of order the Planck scale. It is also
deduced from the solution of Einstein's field equations that TeV
and Planck branes have equal magnitude but opposite sign tensions
and $\Lambda<0$. Therefore, the spacetime in between TeV and
Planck branes is a slice of an $AdS_5$ geometry. Inserting metric
(\ref{RSmetric}) into the action for the gravity and integrating
over extra dimensional coordinate $y$, we obtain the following
relation between the Planck scale and the fundamental scale:
\begin{eqnarray}
\bar{M}^{2}_{Pl}=\frac{M^{3}}{k}(1-e^{-2kr_{c}\pi}).
\end{eqnarray}
If $k\sim \bar{M}_{Pl}$ and $e^{-2kr_{c}\pi}$ is very small then
the hierarchy between $\bar{M}_{Pl}$ and $M$ is eliminated. From
the action for matter fields we can deduce that any mass scale
$m_0$ on the TeV brane in the higher dimensional theory will
correspond to a physical mass $e^{-kr_{c}\pi}m_0$. The factor
$e^{-kr_{c}\pi}$ is called warp factor. If $kr_{c}\sim 12$ then
the warp factor is small enough to generate TeV scale masses from
the masses of order $M_{Pl}$. Hence, RS model solves the hierarchy
problem without generating a large hierarchy between $k$ and
$\frac{1}{r_c}$.

In the RS model, KK graviton mass spacing is quite large compared
with the ADD model. The mass spectrum is given by
$m_{n}=x_{n}ke^{-kr_{c}\pi}=x_{n}\beta\Lambda_{\pi}$ where
$\beta=k/\bar{M}_{Pl}$ and $x_{n}$ are the roots of $J_{1}(x_{n})=0$
\cite{Davoudiasl:1999jd}. Therefore the mass spacing is at the order
of TeV scale. Summation in the graviton propagator cannot be
approximated to an integral. Instead, discrete graviton mass
spectrum should be considered in the summation. Since the
contribution of the KK graviton excitations to the propagator is
small for masses above the center-of-mass energy of the process, we
can cut off the series at some finite mass value. During
calculations, we have considered first four roots of the Bessel
function. Another important feature of the RS model is that massive
KK graviton excitations couple to SM fields with a coupling constant
$\frac {1}{\Lambda_{\pi}}$ where $\Lambda_{\pi}$ is a scale of the
order of TeV.

Amplitude square for the process $\gamma q\to \gamma q$ in the RS
model can be easily obtained from (\ref{amplitudeKK}) and
(\ref{amplitudeint}) through the replacement
\cite{Davoudiasl:1999jd}:
\begin{eqnarray}
\frac{1}{\bar {M_{Pl}}^2}D(\hat t)\longrightarrow
\frac{1}{2\Lambda_\pi^2}\sum_n\frac{1}{\hat t-m_n^2+im_n\Gamma_n}
\end{eqnarray}
where the decay width for the nth KK graviton excitation is
$\Gamma_{n}=\rho m_{n}(\frac{m_{n}}{\Lambda_{\pi}})^{2}$. Here,
$\rho$ is a constant which is assumed to be 1.

\begin{figure}
\includegraphics[scale=1]{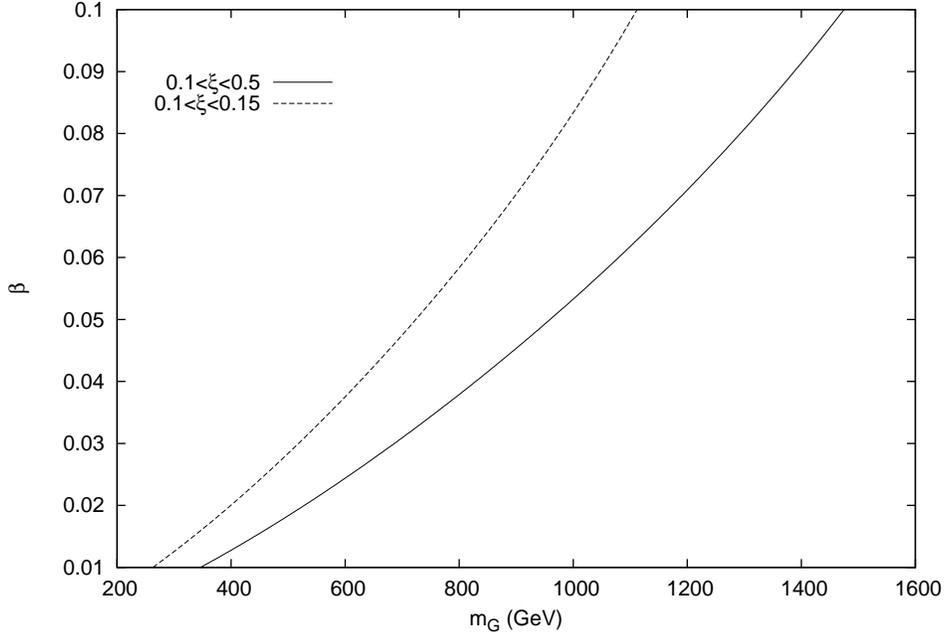}
\caption{Limits in the $\beta-m_G$ plane for an integrated
luminosity of 200$fb^{-1}$. The regions above the curves are
excluded at 95\% C.L. The virtuality of the deep inelastic
scattering is taken to be $Q^2={(5M_Z)}^2$ where $M_Z$ is the mass
of the Z boson.\label{fig5}}
\end{figure}

We have obtained 95\% C.L. excluded regions in the $\beta-m_G$ plane
using a similar statistical analysis that was performed for the ADD
model. Here, $m_G$ is the mass of the first KK graviton excitation,
i.e., $m_G=m_1$. We present our results in Fig.\ref{fig5} for two
different forward detector acceptances $0.1<\xi<0.15$ and
$0.1<\xi<0.5$. The limits for the whole AFP and CMS-TOTEM acceptance
regions are weaker than the limits for $0.1<\xi<0.15$ and
$0.1<\xi<0.5$ subintervals. Hence, we do not present them. In the
ADD model we have examined the validity of Bjorken scaling by
considering three different virtuality values. We have showed that
although the Bjorken scaling is not strictly valid, the limits vary
a little with $Q^2$. We expect the same behavior in the RS model
case since we have used the same distribution functions. Therefore,
we present our limits only for ${(5M_Z)}^2$.

\section{Conclusions}
The potential of $\gamma\gamma$ processes at the LHC to probe large
and warped extra dimensions was investigated in Refs.
\cite{lhc3,lhc5,lhc11} by some of the authors of this work. In these
earlier papers the processes $pp\to p\gamma\gamma p\to p \ell^+
\ell^- p$ \cite{lhc3}, $pp\to p\gamma\gamma p\to p \gamma \gamma p$
\cite{lhc5} and $pp\to p\gamma\gamma p\to p t \bar t p$ \cite{lhc11}
were considered and the bounds on model parameters of ADD and RS
models were obtained. In our present paper we have probed the large
and warped extra dimensions via $\gamma q\to \gamma q$ subprocess in
a {\it quasireal photon-proton} deep inelastic scattering at the
LHC. In the case of ADD model, the bounds that we have obtained for
the $\gamma q\to \gamma q$ subprocess are better than the bounds
obtained in our earlier papers. For instance, in the acceptance
region of $0.1<\xi<0.5$ and $L_{int}=100 fb^{-1}$ the lower bounds
on the cutoff scale $\Lambda_T$ for the subprocesses $\gamma \gamma
\to \ell^+ \ell^-$, $\gamma \gamma \to \gamma \gamma$ and $\gamma
\gamma \to t \bar t$ are 3500 GeV, 5100 GeV and 2700 GeV
respectively. On the other hand, same bound for $\gamma q\to \gamma
q$ subprocess exceeds 6000 GeV. In the case of RS model, excluded
region of the model parameters extends to wider regions than the
case of the subprocesses $\gamma\gamma \to \ell^+ \ell^-$ and
$\gamma \gamma \to t \bar t$. When we compare our present bounds
with the bounds obtained from the subprocess $\gamma\gamma \to
\gamma \gamma$, we see that our present bounds are little better
than the bounds from $\gamma\gamma \to \gamma \gamma$. But the
difference in bounds is minor especially for low $\beta$ values. For
instance, $95\%$ C.L. lower bounds on the mass of the KK graviton
obtained from the subprocess $\gamma\gamma \to \gamma \gamma$ are
910 GeV and 1350 GeV for $\beta=0.05$ and $\beta=0.1$ respectively.
Here, the forward detector acceptance is taken to be $0.1<\xi<0.5$
and $L_{int}=200 fb^{-1}$. The same bounds for $\gamma q\to \gamma
q$ subprocess are 965 GeV and 1466 GeV for $\beta=0.05$ and
$\beta=0.1$ respectively. Hence, we can say that the bounds obtained
from the subprocess $\gamma q\to \gamma q$ are comparable to those
obtained from $\gamma\gamma \to \gamma \gamma$.

The reason why the subprocess $\gamma q\to \gamma q$ provides more
stringent bounds than the above $\gamma\gamma$ processes, is a
consequence of a fact related to quark and photon distributions. In
general, quark distribution functions are bigger in magnitude than
equivalent photon distribution functions, i.e., in a proton the
probability to find a quark with a Bjorken parameter $x$ is higher
than the probability of a quasireal photon with same momentum
fraction $x$. Therefore, $\gamma$-quark processes have higher
effective luminosity than $\gamma\gamma$ processes. Furthermore,
although both quark and quasireal photon distributions decrease as
the $x$ parameter increases, this behavior is drastic in the
quasireal photon case. Thus, quarks in general carry more of the
proton's energy than quasireal photons. Hence, $\gamma$-quark
processes have higher energy reach than $\gamma\gamma$ processes.
Due to above reasons, we expect that $\gamma$-quark processes have a
higher potential in probing new physics compared with $\gamma\gamma$
processes.

The subprocess $\gamma q\to \gamma q$ in the $\gamma$-proton
collision seems to have lower potential in probing RS model than ADD
model of extra dimensions. The reason for this feature is related to
the fact that KK graviton contributions in $\gamma q\to \gamma q$
take part only in a t-channel exchange diagram. A detailed
explanation can be given as follows: In the ADD model, graviton mass
spacing in a KK tower is very narrow. Hence, it is assumed that the
mass of the KK states is continuously distributed. It is impossible
to see the effect of an individual ADD graviton but their cumulative
effect might be observable. On the other hand, in the RS model KK
graviton mass spacing is quite large, $\sim {\cal O}{(TeV)}$. At the
LHC energies we hope to discover the first KK excitation which has a
mass of order 1 TeV. Due to this discrete mass structure of RS
gravitons resonance effects are important in the RS model but they
are absent in the case of ADD model. It is obvious that resonance
effects are not observed in u or t-channel exchange diagrams.
Instead, they appear in processes including s-channel graviton
exchange diagrams. For this reason, our bounds on RS model
parameters are only slightly better than the bounds from
$\gamma\gamma \to \gamma \gamma$ although $\gamma$-quark processes
have higher energy reach and effective luminosity with respect to
$\gamma\gamma$ processes.

Recent results on large and warped extra dimensions from CMS and
ATLAS experiments provide stringent limits
\cite{Marionneau:2013fna,ADD1,ADD2,ADD3,Aad:2012bsa,Aad:2012cy}. In
the case of ADD model, our limits on $\Lambda_T$ for an acceptance
of $0.1<\xi<0.5$ are better than these current experimental bounds.
However, our limits on RS model parameters are weaker than the
recent experimental bounds \cite{Aad:2012cy}. Therefore the reaction
$pp\to p\gamma p\to p\gamma qX$ has a considerable potential in
probing large extra dimensions of the ADD model. On the other hand,
its potential is relatively low for the case of RS model.

\begin{acknowledgments}
This work has been supported by the Scientific and Technological
Research Council of Turkey (T\"{U}B\.{I}TAK) in the framework of
the project no: 112T085.
\end{acknowledgments}

\appendix*
\section{Equivalent photon approximation and photon spectrum}
Incoming photon beam in the subprocess $\gamma q\to \gamma q$ is
described by EPA. According to EPA, equivalent photon distribution
for photons which are emitted from a proton beam is given through
the formula \cite{budnev1975,baur2002,piotrzkowski2001}:
\begin{eqnarray}
\label{spectrum1}
\frac{dN_\gamma}{dE_{\gamma}dQ^{2}}=\frac{\alpha}{\pi}\frac{1}{E_{\gamma}Q^{2}}
[(1-\frac{E_{\gamma}}{E})
(1-\frac{Q^{2}_{min}}{Q^{2}})F_{E}+\frac{E^{2}_{\gamma}}{2E^{2}}F_{M}]
\end{eqnarray}
where
\begin{eqnarray}
&&Q^{2}_{min}=\frac{m^{2}_{p}E^{2}_{\gamma}}{E(E-E_{\gamma})},
\;\;\;\; F_{E}=\frac{4m^{2}_{p}G^{2}_{E}+Q^{2}G^{2}_{M}}
{4m^{2}_{p}+Q^{2}} \\
G^{2}_{E}=&&\frac{G^{2}_{M}}{\mu^{2}_{p}}=(1+\frac{Q^{2}}{Q^{2}_{0}})^{-4},
\;\;\; F_{M}=G^{2}_{M}, \;\;\; Q^{2}_{0}=0.71 \mbox{GeV}^{2}
\end{eqnarray}
In the above formula, $Q^2$ and $E_\gamma$ are the virtuality and
energy of the photon spectrum. E is the energy of the incoming
proton beam. $m_{p}$ and $\mu_{p}$ denote the mass and the magnetic
moment of the proton. $F_{E}$ and $F_{M}$ are functions of the
electric and magnetic form factors. After integration over $dQ^2$ in
the interval $Q^{2}_{min}-Q^{2}_{max}$, equivalent photon
distribution can be written as \cite{kepka2008}
\begin{eqnarray}
\label{spectrum2} \frac{dN_\gamma}{dE_{\gamma}}=\frac{\alpha}{\pi
E_{\gamma}}
\left(1-\frac{E_{\gamma}}{E}\right)\left[\varphi\left(\frac{Q^{2}_{max}}{Q^{2}_0}\right)
-\varphi\left(\frac{Q^{2}_{min}}{Q^{2}_0}\right)\right].
\end{eqnarray}
Here, the function $\varphi$ is defined by
\begin{eqnarray}
\varphi(x)=(1+ay)\left[-ln(1+\frac{1}{x})+\sum_{k=1}^3\frac{1}{k(1+x)^k}\right]+\frac{y(1-b)}{4x(1+x)^3}\nonumber\\
+c\left(1+\frac{y}{4}\right)\left[ln\left(\frac{1-b+x}{1+x}\right)+\sum_{k=1}^3\frac{b^k}{k(1+x)^k}\right]
\end{eqnarray}
where
\begin{eqnarray}
y=\frac{E_\gamma^2}{E(E-E_\gamma)},\;\;\;\;\;\;a=\frac{1+\mu^{2}_{p}}{4}+\frac{4m^2_{p}}{Q^{2}_0}\approx7.16 \nonumber\\
b=1-\frac{4m^2_{p}}{Q^{2}_0}\approx-3.96,\;\;\;\;\;\;c=\frac{\mu^{2}_{p}-1}{b^4}\approx0.028
\end{eqnarray}
 The contribution to the integral above
$Q^{2}_{max}\approx2\;GeV^2$ is negligible. Therefore during
calculations we set $Q^{2}_{max}=2\;GeV^2$.

\newpage

\end{document}